\documentstyle[twoside,fleqn,espcrc2]{article}

\newcommand{\be}{\begin{equation}}
\newcommand{\ee}{\end{equation}}
\newcommand{\bd}{\begin{displaymath}}
\newcommand{\ed}{\end{displaymath}}
\newcommand{\bea}{\begin{eqnarray}}
\newcommand{\eea}{\end{eqnarray}}
\newcommand{\x}{\mbox{x}}

\newcommand{\AmS}{{\protect\the\textfont2
  A\kern-.1667em\lower.5ex\hbox{M}\kern-.125emS}}

\title{DIS Structure Functions in Lattice QCD\thanks{Work supported in part 
       by the EC Contracts
     ``Computational Particle Physics'' ERBCHBGCT940665 and 
       ERBCHRXCT920051.}$^,$\thanks{Presented by Stefano Capitani.}
}

\author{Stefano Capitani\address{University of Southampton, Dept. of Physics, 
        Highfield, Southampton SO17 1BJ, United Kingdom}$^,$\thanks{From 
        October 1996 at the Desy Theory Group.}
        $\!$and Giancarlo Rossi\address{Dipartimento di Fisica, Universit\`a 
        di Roma ``Tor Vergata'' and INFN, Sezione di Roma II, \\
        Via della Ricerca Scientifica 1, I-00133 Roma - Italy}}

\begin{document}

%SHEP 96/boh \hfill \hfill \hfill \hfill August boh, 1996

\begin{abstract}
In this talk I present the complete 1-loop perturbative computation 
of the renormalization constants and mixing coefficients of quark and
gluon lattice operators of rank two and three whose hadronic elements
enter in the determination of the first and second moment of Deep
Inelastic Scattering Structure Functions, making use of the
nearest-neighbor improved ``clover-leaf'' lattice QCD action.

To perform the huge amount of calculations required for the evaluation
of all the relevant Feynman diagrams, extensive use of symbolic 
manipulation languages like Schoonschip and Form has been made.
\end{abstract}

% typeset front matter (including abstract)
\maketitle

\section{Introduction}

The computation on the lattice of renormalization factors is a necessary 
ingredient to connect lattice operators and matrix elements
to their continuum counterparts and extract physical 
quantities from Monte Carlo results.
We report here on the complete computation of the renormalization constants 
and mixing coefficients of the quark and gluon operators of rank two and of 
the quark operators of rank three~\cite{capitani} that measure the first two 
moments of Deep Inelastic Scattering Structure Functions (SF's). 
We have computed all these constants using the Sheikholeslami-\-Wohlert (SW) 
$O(a)$ improved action with $c_{SW}=1$ in 1-loop perturbation theory. 
Improvement~\cite{Sym,Lus,She} reduces the systematic error 
associated with the finiteness of the lattice spacing $a$, which is lowered, 
for on-shell quantities, from $O(a)$ to $O(a/\log a)$.

To work out the huge expressions arising for each of
the numerous Feynman diagrams, it has been necessary to use extensively the 
algebraic computer manipulation languages Schoonschip and Form.

\section{Moments of Structure Functions}

The hadronic tensor $W_{\mu \nu}$, from which the SF's are
defined, is given in terms of the hadronic currents by
\be
W_{\mu \nu} = \frac{1}{2 \pi} \int\! d^4 x \: e^{\displaystyle i q x} 
\langle p | J_{\mu}(x) J_{\nu}(0) | p \rangle. 
\ee 
Using the Wilson OPE near the light-like region, $A(x) B(0) \sim \sum_{N,i} 
c_{N,i}(x^{2}) x^{\mu_1} \cdots x^{\mu_N} O^{(N,i)}_{\mu_{1} \cdots 
\mu_{N}}(0)$, the local product of operators can be expressed in terms of a 
set of symmetric and traceless operators $O^{(N,i)}_{\mu_{1} \cdots \mu_{N}}$ 
with vanishing vacuum expectation values. The matrix elements of the dominant 
(twist two) operators in the expansion, $O^{(N)}_{\mu_{1} \cdots \mu_{N}}$, 
have the general form
\be
\langle p|O^{(N)}_{\mu_{1}\cdots\mu_{N}}|p \rangle =A_N(\mu)~p_{\mu_{1}}\cdots 
p_{\mu_{N}} +~\mbox{traces} \label{eq:pppp}.
\ee
The formula for the moments of the SF's becomes
%\be
%\langle \x^{N-1} \rangle =\int\!d\x\:\x^{N - 1}{\cal F}_{k}(q^{2},\x)
%=C_{N}(\frac{q^{2}}{\mu^2}) A_{N}(\mu) \label{eq:chm},
%\ee
\be
\langle \x^{N-1} \rangle =\!\!\!\int\!d\x \x^{N - 1}{\cal F}_{k}(q^{2},\x)
=C_{N}(\frac{q^{2}}{\mu^2}) A_{N}(\mu) \label{eq:chm},
\ee
where ${\cal F}_{1} = 2 F_{1}, {\cal F}_{2} = F_{2}/\x$ and 
${\cal F}_{3} = F_{3}$. The coefficients $C_N$ are known from continuum 
perturbation theory.
From (\ref{eq:chm}) we can extract $\langle \x^{N-1} \rangle$ knowing the 
corresponding matrix element 
$\langle p|O^{(N)}_{\mu_{1}\cdots\mu_{N}}|p \rangle$. 
The latter contains long distance (non-perturbative) physics, thus
the only viable way to compute the moments of 
SF's is with the use of lattice methods. 

We have considered in our calculations the unpolarized SF's,
and in particular we have computed the renormalization constants and mixing 
coefficients of the operators 
\bd
\begin{array}{l}
O^q_{\mu \nu} = \frac{1}{4} \:\overline{\psi} \:\gamma_{\{ \mu} \!
\stackrel{\displaystyle \leftrightarrow}{D}_{\nu \}} 
\psi~~~\longrightarrow~~~\langle \x \rangle_q \\
O^g_{\mu \nu} = \sum_{\rho} \mbox{Tr} \left[ F_{\{\mu \rho} 
F_{\rho \nu\}} \right]~~~\longrightarrow~~~\langle \x \rangle_g \\
O^q_{\mu \nu \tau} = \frac{1}{8} \,\overline{\psi} \,\gamma_{\{ \mu} \!
\stackrel{\displaystyle \leftrightarrow}{D}_{\nu} 
\stackrel{\displaystyle \leftrightarrow}{D}_{\tau \}} 
\psi~~~\longrightarrow~~~\langle \x^2 \rangle_q ,
\end{array}
\ed
where $D_{\mu}$ is the covariant derivative.

\section{Sheikholeslami-Wohlert improvement}

The matrix elements appearing in (\ref{eq:pppp}) need the computation of 
two- and three-point correlation functions~\cite{old}. To reduce discretization
errors the SW improved action is used~\cite{She}. Adding to the Wilson action
the ``clover-leaf'' SW term
\be 
\Delta S^{f}_{I} = - i g_0 a^{4} \sum_{n,\mu \nu} \frac{r}{4 a} \: 
\overline{\psi}_{n}
\sigma_{\mu \nu} F_{n, \mu \nu} \psi_{n} \label{eq:impr} 
\ee
and performing on the spinor fields in the operators $O^q_{\mu \nu}$ and 
$O^q_{\mu \nu \tau}$ the ``rotations'' 
\bd
\psi \longrightarrow \left( 1 - \frac{a r}{2} \stackrel{\displaystyle
\rightarrow }{\not\!\!{D}} \right)\psi~,~\overline{\psi} \longrightarrow 
\overline{\psi} \left( 1 + \frac{a r}{2} 
\stackrel{\displaystyle \leftarrow }{\not\!\!{D}} \right) ,
\ed
leads, in on-shell matrix elements~\cite{Lus}, to a cancellation of all terms 
that in the continuum limit ($g_0^2 \sim 1/\log a$) are effectively 
of order ``$a$''~\cite{She,rom}. 

The systematic error related to the lattice discretization drops in this way 
from order $a$ to order $a/\log a$. Numerically this results in a substantial 
improvement. In fact, while the magnitude of the order $a$ terms is about 
20--30 \%, the magnitude of the order $a/\log a$ terms turns out to be about 
5--10 \%.

\section{Renormalization constants} 

The renormalization constants connect the bare lattice operators, $O(a)$, 
to finite operators, $\widehat{O}(\mu)$, renormalized at a scale $\mu$:
\be
\widehat{O}^l(\mu) = Z_{lk}(\mu a) O^k(a) .
\ee
In the flavor Singlet case there is a mixing between quark and 
gluon operators of the same rank that have the same conserved 
quantum numbers. We then write: 
\be
\begin{array}{c}
\widehat{O}^q = Z_{qq} O^q + Z_{qg} O^g \\
\widehat{O}^g = Z_{gq} O^q + Z_{gg} O^g ,
\end{array} 
\ee
and in this case all elements of the mixing matrix
\be
\left( \begin{array}{cc}
\langle q| O^q |q \rangle & \langle g,\sigma| O^q |g,\sigma \rangle \\
\langle q| O^g |q \rangle & \langle g,\sigma| O^g |g,\sigma \rangle    
\end{array} \right) 
\ee 
have to be computed.

To this mixing (already present in continuum QCD) the lattice regularization adds 
additional mixings, induced by the breaking of (Euclidean) Lorentz 
invariance~\cite{mix}. It is possible, by a careful choice of the Lorentz 
indices, to simplify at least partially the lattice mixing pattern. However,
the higher the moment the more complicated the mixing 
pattern becomes~\cite{capitani,new}. 
In particular, the non-Singlet operators $O^q_{\{12\}}$ and $O^q_{\{123\}}$
are multiplicatively renormalizable on the lattice in the quenched 
approximation, but another useful second moment operator, $O^q_{\rm{DIS}} 
\equiv O^q_{\{411\}} - \linebreak 
- \frac{1}{2} (O^q_{\{422\}} + O^q_{\{433\}})$, is not.

The breaking of Lorentz invariance forced us to develop special 
computer routines to correctly perform the Dirac algebra on the lattice.
They play a key role in our codes which are designed to 
automatically carry out all the steps of the necessary algebraic manipulations,
starting from the elementary building blocks of each Feynman diagram.

\section{Some results}

Putting $Z=1+g_0^2/16\pi^2(\gamma\log a \mu + B)$, one gets for the $B$'s and 
$\gamma$'s of the first moment operators the results of Tables 1 and 2.
\begin{table}
\begin{tabular}{|c|r|c|r|}
\hline  \multicolumn{2}{|c}{Wilson} & \multicolumn{2}{|c|}{Improved} \\
\hline $B^W_{qq}$ & -3.165 & $B^I_{qq}$ & -15.816 \\
$B^W_{qg}$ & 0.208 & $B^I_{qg}$ & -0.743 \\
$B^W_{gq}$ & -5.817 & $B^I_{gq}$ & -4.044 \\
$(B^f_{gg})^W$ & -2.168 & $(B^f_{gg})^I$ & -6.084 \\
\hline
\multicolumn{4}{|c|}{$B_{gg} =$ -15.585} \\
\hline
\end{tabular}
\caption{The lattice constants $B$ for the first moment; $B^f_{gg}$ 
corresponds to the quark loop contribution to the gluon propagator.}
\end{table}
\begin{table}
\begin{tabular}{|c|r|r||c|r|}
\hline & sails & -4 & \multicolumn{2}{|c|}{} \\
$B_{qq}$ & vertex & 5$/$9 & \multicolumn{2}{|c|}{ANOMALOUS} \\
& self-energy & -1 & \multicolumn{2}{|c|}{DIMENSIONS} \\
\cline{2-5} & \multicolumn{1}{|c|}{total} & - 34$/$9 & $\gamma_{qq}$ & 16$/$3 \\
\hline \multicolumn{2}{|c}{$B_{qg}$} & \multicolumn{1}{|r||}{- 4$/$9}
& $\gamma_{qg}$ & 4$/$3 \\
\hline \multicolumn{2}{|c}{$B_{gq}$} & \multicolumn{1}{|r||}{- 22$/$9} 
& $\gamma_{gq}$ & 16$/$3 \\
\hline \multicolumn{2}{|c}{$B_{gg}$} & \multicolumn{1}{|r||}{- 4$/$3} 
& $\gamma_{gg}$ & 0 \\
\hline \multicolumn{2}{|c}{$B^f_{gg}$} & \multicolumn{1}{|r||}{- 10$/$9} 
& $\gamma^f_{gg}$ & 4$/$3 \\
\hline
\end{tabular}
\caption{The continuum constants $B$ in the $\overline{MS}$ scheme and
the anomalous dimensions for the first moment.}
\end{table}
The numbers reported here for the $B$'s are for $r=1$ and differ slightly from
the ones published in the first paper of~\cite{capitani}. They have been 
re-computed by us and cross-checked against the works in~\cite{pelisset}.

Simulations have been performed in the past with the unimproved 
Wilson action~\cite{old,new}. Within errors, Monte Carlo
results are consistent with experiment. No simulation has been as yet performed
in the improved case. Values of the renormalization constants 
in the improved theory are somewhat larger than in the standard Wilson case. 
As an example, we give here a selection of the results for quark operators
(for the complete results see~\cite{capitani}). 
At $\beta =6.0$ and for $r=1$ one gets~\footnote{We have 
decomposed $O^q_{\rm{DIS}}=\frac{1}{3} (O_A+O_B)$ into the non-symmetric 
$O_A \equiv O^q_{411} - \frac{1}{2} (O^q_{422} + O^q_{433})$ and \linebreak
$O_B \equiv O^q_{141} + O^q_{114} - \frac{1}{2} (O^q_{242} + O^q_{224} 
+ O^q_{343} + O^q_{334})$.}
\bd
\begin{array}{l}
(\widehat{O}^q_{\{12\}})^{\rm{W}} = 1.027~(O^q_{\{12\}})^{\rm{W}} \\
(\widehat{O}^q_{\{12\}})^{\rm{I}} = 1.134~(O^q_{\{12\}})^{\rm{I}} \\
(\widehat{O}^q_{\{123\}})^{\rm{W}} = 1.160~(O^q_{\{123\}})^{\rm{W}} \\
(\widehat{O}^q_{\{123\}})^{\rm{I}} = 1.252~(O^q_{\{123\}})^{\rm{I}} \\
(\widehat{O}^q_{\rm{DIS}})^{\rm{W}} = \frac{1}{3} \left[  
1.184~(O_{A})^{\rm{W}} + 1.156~(O_{B})^{\rm{W}} \right] \\
(\widehat{O}^q_{\rm{DIS}})^{\rm{I}} = \frac{1}{3} \left[ 
1.331~(O_{A})^{\rm{I}} + 1.187~(O_{B})^{\rm{I}} \right] ,
\end{array}
\ed
where the first of each pair shows the Wilson results (for which
we are in agreement with~\cite{new}) and the second the improved results.

Finally, we want to mention that results now exist also for the 
polarized SF's, though limited to the Wilson case~\cite{new}.

We thank C. McNeile, A. Pelissetto and H. Perlt for discussions and
for correspondence.

\end{document}